\documentclass[journal=jpclcd, manuscript=letter]{achemso}

\usepackage[utf8]{inputenc}
\usepackage[T1]{fontenc}
\usepackage{mathptmx}
\usepackage{mathtools}
\usepackage{hyperref}
\usepackage[inline]{enumitem}

\usepackage{graphicx}
\usepackage{physics}
\usepackage{makecell}
\usepackage{xcolor}
\usepackage{braket}
\usepackage{amsmath}
\usepackage[T1]{fontenc} 
\newcommand{\mobilityunit}
    {$\textrm{cm}^2 / \textrm{V} \cdot \textrm{s}$}
    
\newcommand{\revmod}[1]{\textcolor{black}{#1}}
\newcommand{\revrevmod}[1]{\textcolor{black}{#1}}

\title{Finite Temperature TD-DMRG for the Carrier Mobility of Organic Semiconductors}

\author{Weitang Li}
 \affiliation{MOE Key Laboratory of Organic OptoElectronics and Molecular
 Engineering, Department of Chemistry, Tsinghua University, Beijing 100084,
 People's Republic of China }
 
\author{Jiajun Ren}
 \affiliation{MOE Key Laboratory of Organic OptoElectronics and Molecular
 Engineering, Department of Chemistry, Tsinghua University, Beijing 100084, People's Republic of China }

\author{Zhigang Shuai}%
\email{zgshuai@tsinghua.edu.cn}
 \affiliation{MOE Key Laboratory of Organic OptoElectronics and Molecular
 Engineering, Department of Chemistry, Tsinghua University, Beijing 100084,
 People's Republic of China }

\begin{document}

\begin{tocentry}
      \includegraphics[width=\textwidth]{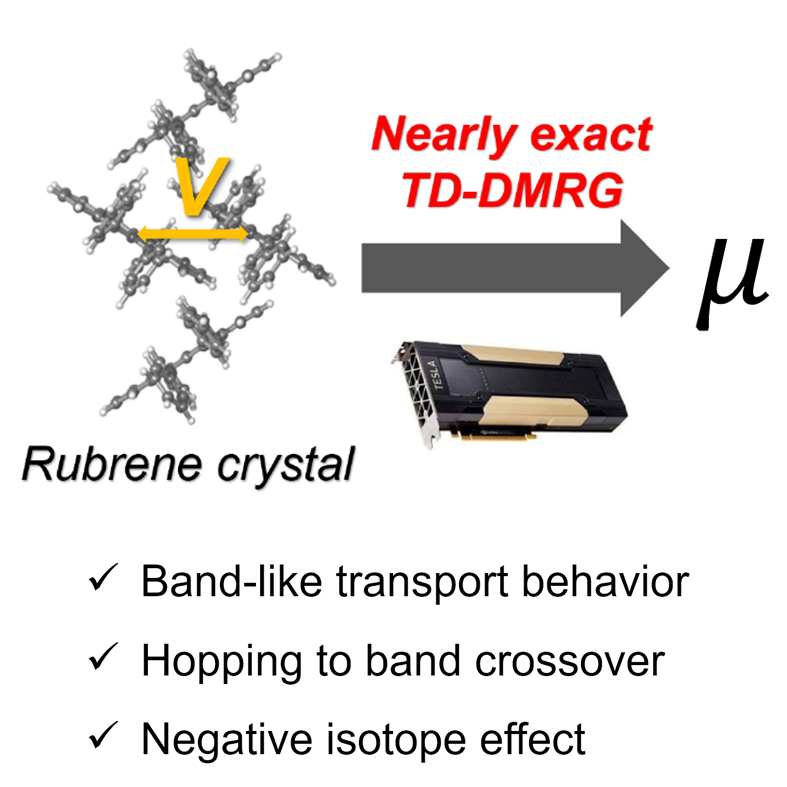}
\end{tocentry}

\begin{abstract}
A large number of non-adiabatic dynamical studies have been applied to reveal the nature of carrier transport in organic semiconductors with different approximations. We present here a ``nearly exact'' graphical process unit (GPU) based finite temperature TD-DMRG method to evaluate the carrier mobility in organic semiconductors as described by electron-phonon model, in particular to rubrene crystal, one of the prototypical organic semiconductors, with parameters derived from first-principles. We find that (i) TD-DMRG is a general and robust method that can bridge the gap between hopping and band picture covering a wide range of electronic coupling strength; and (ii) with realistic parameters, TD-DMRG is able to account for the experimentally observed ``band-like'' transport behavior ($\partial \mu / \partial T < 0$) in rubrene. We further study the long-standing puzzle of isotope effect for charge transport and unambiguously demonstrate that the negative isotope effect ($\partial \mu / \partial m < 0$ where $m$ is the atomic mass) should be universal.
\end{abstract}

\maketitle

Organic semiconductors have attracted strong research interest in the scientific community
due to their potential application in the next generation of electronic
devices~\cite{Forrest04}. Charge mobility is a crucial
physical parameter for device performance.
Qualitative and quantitative descriptions of the charge transport mechanism are essential for molecular design of high efficiency organic electronic device~\cite{bredas07}.
Unfortunately, understanding charge transport mechanism in a microscopic point of view
is of continuous debates due to many-body electron-phonon interaction~\cite{Kenkre82, Mahan00, Blumberger17}.
In the limiting cases of strong or weak electron-phonon interaction,
charge transport behavior can be characterized by 
localized hopping picture~\cite{Holstein2, NAN09} 
or delocalized band transport picture~\cite{GLARUM63, XI12} respectively.
But a large class of organic semiconductors lie in the intermediate regime
where the strength of electronic coupling and electron-phonon interaction is comparable~\cite{Troisi11rev}.
Although a significant amount of effort has been devoted to developing
carrier mobility theory applicable in 
the intermediate regime~\cite{Silbey80, Kenkre89, Kenkre03, Hann041, Hann09, Fratini16, Fratini17},
certain levels of approximation are inevitably involved 
and their effect on the resulting mobility is usually unclear.
For example, 
the approximation adopted by Hannewald and co-workers
is to replace the complicated polaron-coupling operator after polaron transformation
by its thermal average over the phononic part~\cite{Hann09},
\revmod{and in the transient localization theory relaxation time approximation is necessary to connect
carrier mobility with the correlation function of the static tight-binding model~\cite{Fratini16, Fratini17} }.

In the absence of an analytical solution, a number of numerical methods have been applied 
to the carrier mobility problem~\cite{Blumberger17}.
The family of the methods includes
Ehrenfest dynamics~\cite{TROISI06, Fratini11},
surface hopping~\cite{Wang13SH, Blumberger19},
time-dependent wave packet diffusion (TDWPD)~\cite{Zhao13, Cao14},
and so on~\cite{Fratini09, Cao13, Shi15, Nenad19}. 
However, most of the methods have to sacrifice accuracy for applicability to realistic materials.
For example, Ehrenfest dynamics and surface hopping treat the nuclear degree of freedom classically,
and TDWPD performs truncation to the stochastic Schrödinger equation.
It is only until recently that numerically exact methods such as 
hierarchical equations of motion (HEOM)~\cite{Shi10, Shi19}
and quantum Monte-Carlo (QMC)~\cite{Mishchenko15}
are used to tackle the charge transport problem, 
but these explorations are limited to model systems instead of realistic materials. 

Over the past years, density matrix renormalization group (DMRG)
and its time-dependent variant (TD-DMRG) have become powerful numerically
``nearly exact'' solvers
for quantum many-body systems~\cite{white92, jeckelmann1998density, Schol11, Scholl19}.
The application of TD-DMRG to electron-phonon correlated systems
has also been a great success~\cite{prior2010efficient, Ren18, Ma19, Reiher19},
achieving accuracy comparable to multi-configuration time-dependent Hartree (MCTDH), 
the de facto state-of-the-art method for non-adiabatic dynamics in complex system~\cite{beck2000multiconfiguration}.
The latter is rarely applied for finite temperature due to the computational burden.
Recently, we have implemented a highly efficient GPU-based finite-temperature TD-DMRG with projector splitting algorithm for time evolution, which can tremendously increase the computational power while keeping the “nearly exact” nature~\cite{Ren18, li2020numerical}.
Now we apply this method to evaluate carrier mobility $\mu$ in rubrene crystal
which is frequently studied as one of the prototypical organic semiconductor materials~\cite{Ortmann17}.
Our calculation based on first-principle parameters predicts ``band-like" transport behavior ($\pdv{\mu}{T} < 0$) 
for rubrene crystal, which is in agreement with experimental observations~\cite{Podzorov05, Machida10, Ren17}.
Additionally, we find that if the electronic coupling is set to be an adjustable parameter,
TD-DMRG is able to bridge the gap between hopping and band picture by
reproducing the analytical formula in hopping limit
and approaching the asymptotic behavior in the band limit.
We further apply the method to study the long-standing puzzle of
isotope effect for charge transport
pioneered by Munn \textit{et al.} in 1970~\cite{Munn70},
and confirm the negative isotope effect~\cite{Jiang18}
($\partial \mu / \partial m < 0$ where $m$ is the atomic mass)
caused by reduced polaron size.

Theoretical investigations have revealed that in the rubrene crystal 
the electronic coupling \revmod{and the hole mobility (predicted by the TDWPD method)} at the stacking direction is much larger than other
directions~\cite{TROISI07, Jiang16, Troisi18},
so we map the rubrene crystal to a one-dimensional multi-mode Holstein model.
\revrevmod{The one-dimension approximation is commonly adopted by researchers when new methodologies to compute mobility are proposed\cite{TROISI07, Fratini12, Wang13SH}.
The Hamiltonian of the model reads:}
\begin{equation}
\label{eq:ham}
\begin{aligned}
    \hat H & = \hat H_{e} + \hat H_{ph} + \hat H_{e-ph} \\ 
    \hat H_e & = V \sum_n (c_{n+1}^\dagger c_n +  c_{n}^\dagger c_{n+1}) \\
    \hat H_{ph} & = \sum_{n, m} \omega_m b^\dagger_{n, m} b_{n, m} \\
    \hat H_{e-ph} & = \sum_{n, m} g_m \omega_m (b^\dagger_{n, m} + b_{n, m}) c^\dagger_n c_n
\end{aligned}
\end{equation}
where $c^\dagger$ ($c$) and $b^\dagger$ ($b$) are the creation (annihilation) operator
for electron and phonon respectively, 
$V$ is the electronic coupling (also known as transfer integral),
$\omega_m$ is the frequency of the $m$th normal mode and
$g_m$ is the dimensionless electron-phonon coupling constant between the $m$th mode with the electronic degree of freedom.
In this paper we use $V=83$ meV and 
9 vibration modes for each molecule with
vibration frequency ranging from $84 \ \textrm{cm}^{-1}$
to $1594 \ \textrm{cm}^{-1}$ and
the total reorganization energy
$\lambda = \sum_m \lambda_m = \sum_m g^2_m\omega_m =75$ meV 
unless otherwise stated.
The parameters are adopted from our previous work~\cite{Jiang16}.
The $\omega_m$ and $\lambda_m$ for each vibration modes 
are shown in the supporting information.
\revrevmod{It should be noted that in a number of publications off-diagonal electron-phonon coupling  (also known as Peierls coupling or dynamic disorder) is considered to be dominant in rubrene crystal~\cite{TROISI07, Fratini17}, however, the conclusion is drawn from quite approximate methodologies such as Ehrenfest dynamics and 
should be subjected to verification by higher-level methods.
At the end of this paper we demonstrate TD-DMRG can be generalized to models with off-diagonal electron-phonon coupling and we leave thorough investigation
on the controversial role of off-diagonal electron-phonon coupling in organic semiconductors~\cite{Shi10, Wang10Multicale, Wang13SH, Shi15} to future work.}

\begin{figure}
  \includegraphics[width=.48\textwidth]{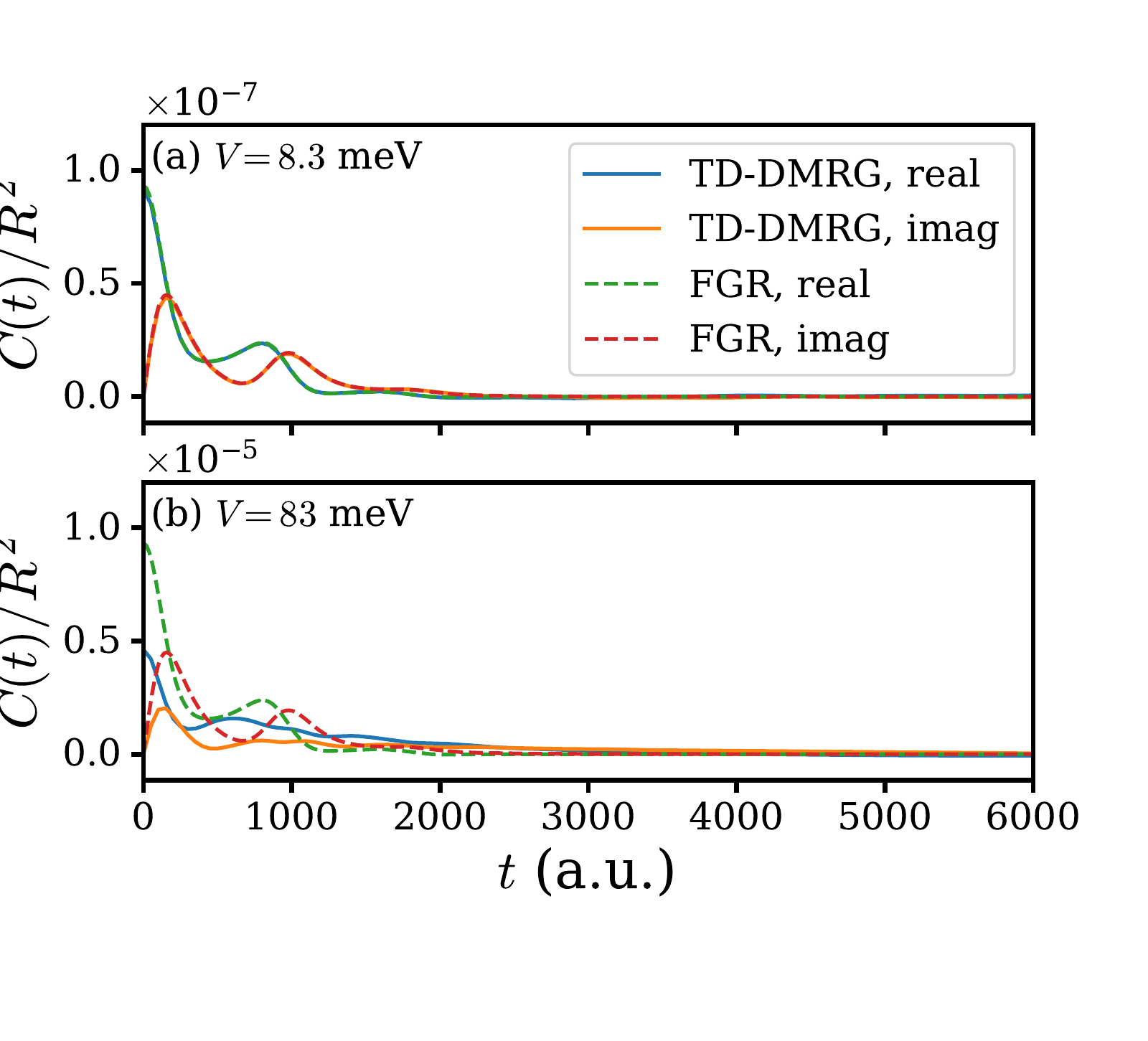}
  \caption{Comparison of the correlation function $C(t)/R^2$ obtained by TD-DMRG (solid) and analytical formula Eq.\ref{eq:ct} derived from FGR valid at hopping limit (dashed).
  (a) $V$ is set to 8.3 meV where analytical solution is available in order to demonstrate the accuracy of the TD-DMRG method. 
  (b) $V$ is set to 83 meV which is the actual parameter for the rubrene crystal. $C(t)/R^2$ is in atomic unit.}
  \label{fig:ct}
\end{figure}

The carrier mobility is obtained via Kubo formula~\cite{Mahan00}:
\begin{equation}
\label{eq:kubo}
    \mu = \frac{1}{2k_B T e_0} \int_{-\infty}^{\infty} \braket{\hat j(t) \hat j(0)} dt
    = \frac{1}{2k_B T e_0} \int_{-\infty}^{\infty} C(t) dt
\end{equation}
where for the Holstein Hamiltonian:
\begin{equation}
    \hat j = \frac{e_0 V R}{i} \sum_n (c_{n+1}^\dagger c_n  -  c_{n}^\dagger c_{n+1}) .
\end{equation}
Here $R$ is the inter-molecular distance.
\revmod{
With the 9-mode Holstein Hamiltonian for the rubrene crystal, 
$C(t)$ rapidly decays to nearly zero before 6000 a.u. as shown in Figure~\ref{fig:ct}.
We note that $C(t)$ is expected to show some kind of Poincaré recurrence 
because we have treated the rubrene crystal as a closed system. 
However, the recurrence is unlikely to happen for realistic materials due to the presence of various dissipations.
Therefore, when integrating the correlation function 
the integration time limit is set to the nearly zero region before the recurrence time to exclude the effect of the artificial recurrence. An example of the recurrence is included in the supporting information.}

The evaluation of the current-current correlation function $C(t)=\braket{\hat j(t) \hat j(0)}$
is performed by TD-DMRG through imaginary and real time propagation.
The basic idea behind TD-DMRG in the language of matrix product state (MPS)
and matrix product operator (MPO) has already been reviewed in detail~\cite{Schol11, Scholl19} 
\revmod{and a short overview can be found in the supporting information}. 
Here we only briefly summarize the finite temperature algorithm based on thermal field dynamics, 
also known as the purification method~\cite{White05ft,Schol11}.
The thermal equilibrium density matrix of any mixed state in physical space $P$ can be expressed as a partial trace over an enlarged Hilbert space $P\otimes Q$, where $Q$ is an auxiliary space chosen to be a copy of $P$.
The thermal equilibrium density operator is then expressed
as a partial trace of the pure state $\Psi_\beta$ in the enlarged Hilbert space over the $Q$ space:
\begin{equation}
    \hat \rho_\beta = \frac{e^{-\beta \hat{H}}}{Z} = \frac{\Tr_Q\ket{\Psi_{\beta}}\bra{ \Psi_{\beta}}}
    {\Tr_{PQ}\ket{\Psi_{\beta}}\bra{ \Psi_{\beta}}}
\end{equation}
and the pure state $\ket{ \Psi_\beta }$
represented as an MPS is obtained by the imaginary time propagation
\revmod{from the locally maximally entangled state $\ket{I}=\sum_i \ket{i}_P \ket{i}_Q$ to $\beta/2$}:
\begin{equation}
    \ket{\Psi_\beta} = e^{-\beta \hat H / 2} \ket{I} .
\end{equation}

To calculate $C(t)$, 
$\ket{\Psi_\beta}$ and $\hat j (0) \ket{\Psi_\beta}$ 
are propagated in real time to obtain
$e^{-i \hat H t} \ket{\Psi_\beta}$ and $e^{-i \hat H t} \hat j (0) \ket{\Psi_\beta}$ and then $C(t)$ is calculated by:
\begin{equation}
    C(t) = \bra{\Psi_\beta} e^{i \hat H t} \hat j (0) e^{-i \hat H t} \hat j (0) \ket{\Psi_\beta} / Z .
\end{equation}
Here the current operator $\hat j (0)$ is 
represented as an MPO and inner-product for $\ket{\Psi_\beta}$ includes tracing over 
both $P$ space and $Q$ space.
In principle, both imaginary and real time propagation can be carried
out by any time evolution methods available to TD-DMRG~\cite{Scholl19}.
In this work, we use the time-dependent variational principle based projector splitting time evolution scheme~\cite{Haeg16}, which is found to be relatively efficient and accurate in our recent work~\cite{li2020numerical}.
Readers are also referred to
our previous work on finite temperature TD-DMRG~\cite{Ren18} 
for more computational details.
In most of our simulations, 
the number of molecules in the periodic one-dimensional chain is 21 and the bond dimension is 32.
In certain cases, we find a larger system size and bond dimension necessary,
such as at low temperature ($T=200 \ \textrm{K}$) 
or large electronic coupling ($V=500 \ \textrm{meV}$).

\revmod{
Before presenting calculated mobility for the rubrene crystal, 
we show Figure~\ref{fig:ct}a the benchmark for the TD-DMRG method by comparing the correlation function calculated at $V=8.3$ meV
with analytical solution valid in the hopping limit  obtained from Fermi's Golden Rule (FGR)~\cite{NAN09}:
\begin{equation}
\label{eq:ct}
    C(t) / R^2 = V^2 \exp{-\sum_m g_m^2[2n_m + 1 - n_m e^{-i\omega_m t} 
    - (n_m+1)e^{i\omega_m t}]}
\end{equation}
where $n_m$ is the thermal average occupation number of the $m$th vibrational mode
at a given temperature. 
We can see from Figure~\ref{fig:ct}a that in the hopping limit TD-DMRG is able to reproduce the analytical formula with impressive precision.
The two curves by TD-DMRG and Eq.~\ref{eq:ct} coincide exactly with each other
for both real and imaginary part.
We also show in Figure~\ref{fig:ct}b the same comparison
but with $V=83$ meV, the actual parameter of the rubrene crystal which is in the intermediate regime.
For parameters of realistic material Eq.~\ref{eq:ct}
clearly fails and a more accurate method is required. 
}

\begin{figure}
  \includegraphics[width=.48\textwidth]{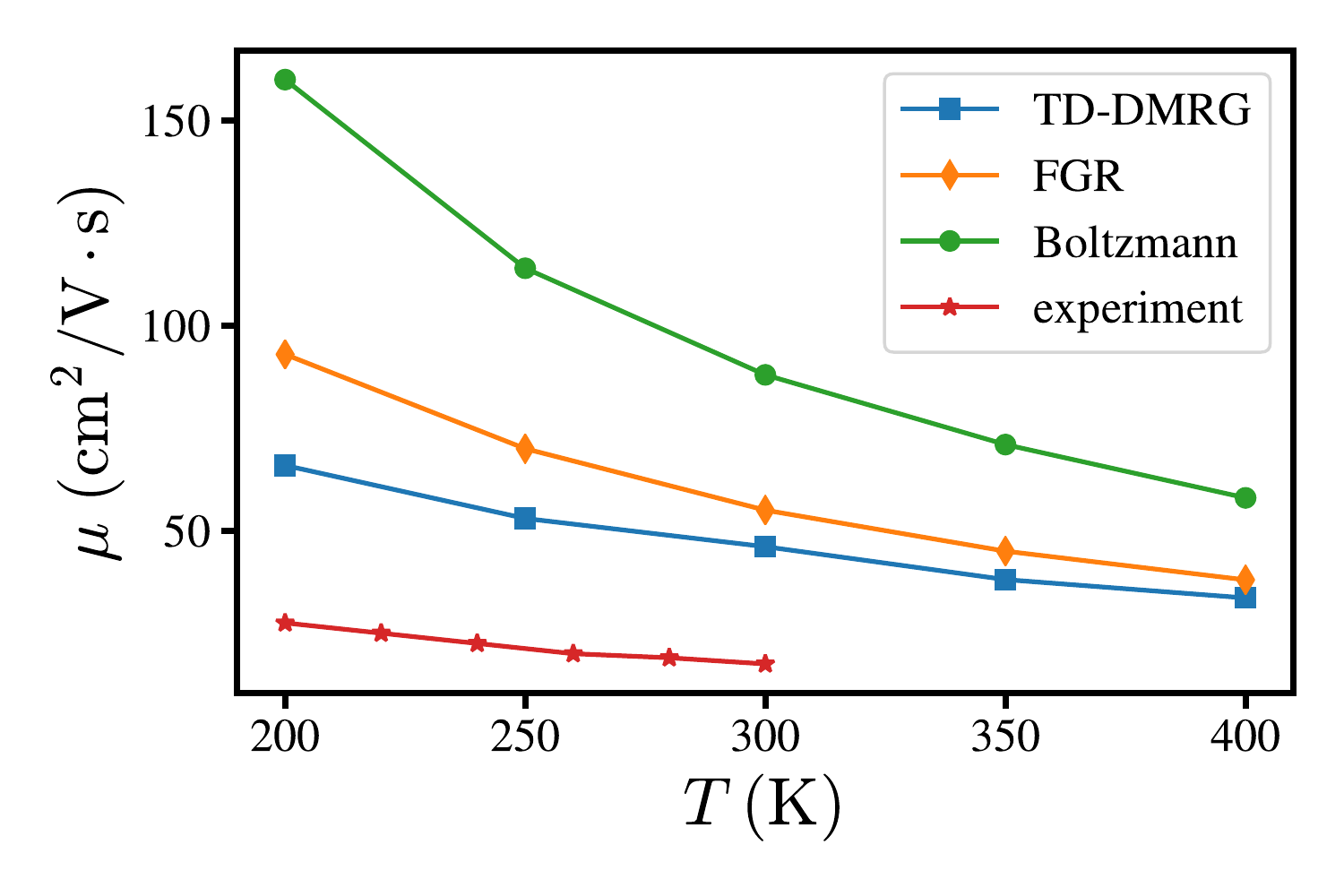}
  \caption{Carrier mobility of rubrene crystal from $T=200$ K to 400 K obtained from TD-DMRG along with experimental results~\cite{Ren17}. The results obtained from FGR and Boltzmann transport theory are also shown for comparison, which are respectively valid at the hopping regime and the band regime. }
  \label{fig:mu-t}
\end{figure}

\begin{figure}
  \includegraphics[width=.48\textwidth]{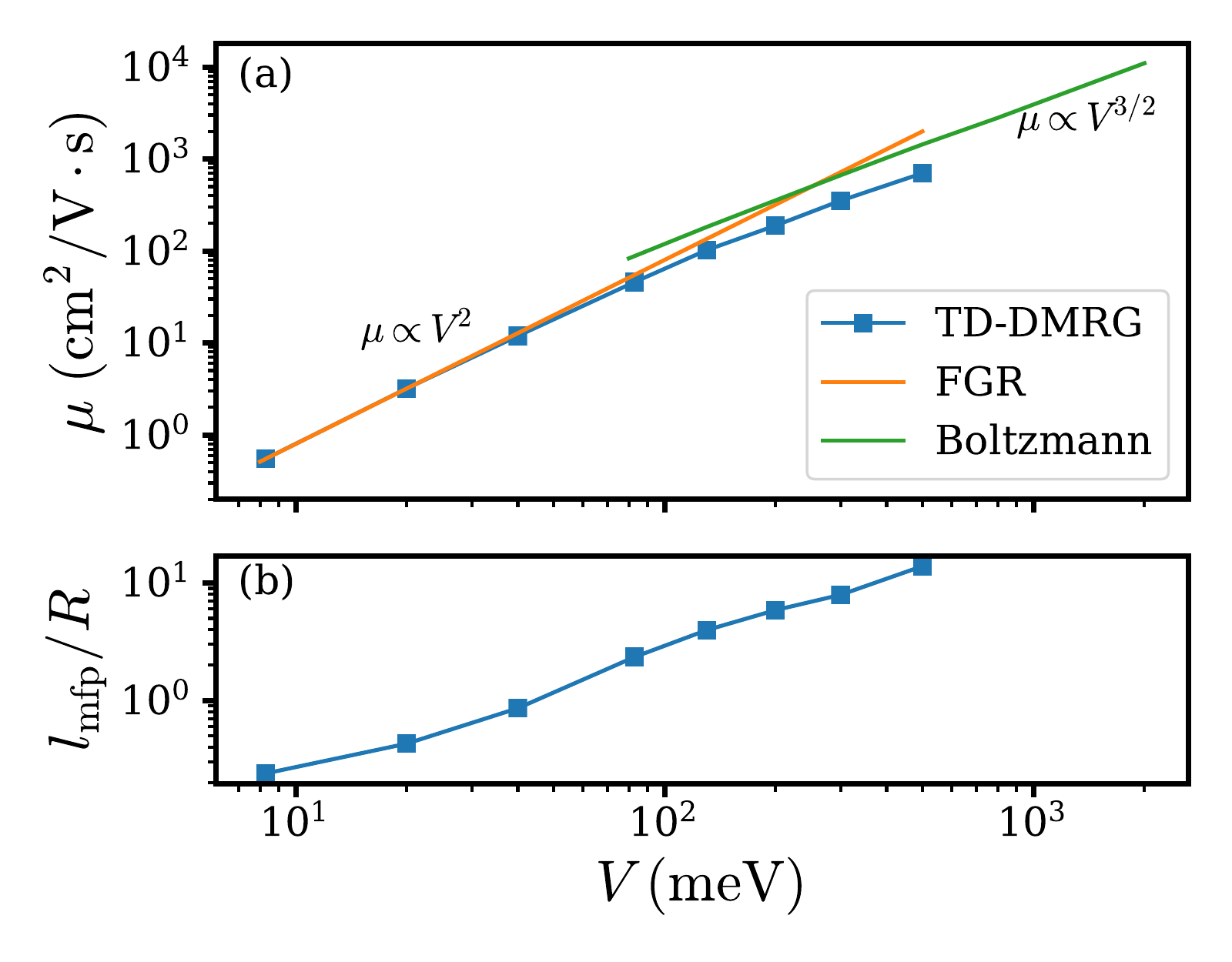}
  \caption{(a) Carrier mobility from weak to strong electronic coupling ($V$) calculated by TD-DMRG, FGR (hopping limit) and Boltzmann transport theory (band limit).
  (b) Mean free path ($l_{\textrm{mfp}}$) of the charge carrier from weak to strong electronic coupling calculated by TD-DMRG. In both panels the electron-phonon coupling parameters are adopted from the first-principle calculation of rubrene crystal.}
  \label{fig:mu-v}
\end{figure}

Figure~\ref{fig:mu-t} shows the carrier mobility of rubrene crystal 
obtained from TD-DMRG at various temperatures
along with results from experiment~\cite{Ren17}.
The ``band-like'' transport behavior that the mobility decreases with temperature is captured by TD-DMRG
and the absolute mobility value (47 \mobilityunit) at $T=300\ \textrm{K}$
closely matches
the highest experimental report (40 \mobilityunit)~\cite{Ogawa07}.
\revrevmod{Nevertheless, the most commonly reported experimental room temperature mobility 
for rubrene crystal is around 15-20 \mobilityunit ~\cite{Sundar04, Bao07, Podzorov04, Ren17}. 
It should be noted that for a number of organic semiconductors including pentacene, the measured mobility has been increased by a factor of about 40 in the past two decades~\cite{Nelson98, Jurchescu04}. A direct comparison with experimental values should also include both static and possibly dynamic off-diagonal disorders.}
For comparison with TD-DMRG results, two widely used methods valid respectively at the incoherent hopping regime and coherent band regime
are also shown in Figure~\ref{fig:mu-t}. 
In the hopping regime, the carrier diffusion is described as site-to-site hopping
with transition rates given by FGR~\cite{NAN09},
whereas the carrier mobility in the band regime is described by the Boltzmann transport theory
with relaxation time determined by the first order perturbation treatment of the electron-phonon interaction.
Although it is natural for Boltzmann transport theory to predict ``band-like''
behavior, the same is not true for the hopping model unless nuclear
tunnelling effect is taken into account, which becomes weaker at
higher temperature, and thus the mobility decreases with temperature~\cite{NAN09}.
Figure~\ref{fig:mu-t} also shows that although both FGR and Boltzmann transport theory 
overestimate carrier mobility, the prediction by FGR is more accurate
than Boltzmann transport theory especially at high temperature.
The observation naturally leads to the question of 
at which strength of electronic coupling does the delocalized band
picture better describes the transport mechanism than the localized hopping picture,
so in Figure~\ref{fig:mu-v}a we further compare TD-DMRG, FGR and
Boltzmann transport theory from weak to strong electronic coupling
while holding the electron-phonon coupling strength constant.
When $V \ll \lambda$ we find TD-DMRG reproduces FGR results perfectly 
\revmod{as expected from the benchmark result in Figure~\ref{fig:ct}a}.
As the electronic coupling increases, the carrier mobility predicted by FGR increases
quadratically while TD-DMRG result increases sub-quadratically,
leading to a noticeable difference between the two methods at $V=83$ meV,
the actual electronic coupling in rubrene crystal.
At this point, the delocalized band picture sets in and shows a
$\mu \propto V^{\frac{3}{2}}$ behavior~\cite{GLARUM63}.
Accordingly, the slope of the TD-DMRG curve gradually 
changes from 2 to $\frac{3}{2}$ and
the carrier mobility by TD-DMRG further deviates from the FGR result.
Unfortunately we can not reproduce band limit as accurately
as hopping limit because of the computationally
prohibitive bond dimension required for
large $V$ and we estimate the data for $V=500$ meV has an error larger than 10\%.
In this regime, FGR predicts a spurious $\mu \propto V^2$ growth
and finally yields mobility higher than the band description at $V > 200 \  \textrm{meV}$,
indicating the complete failure of the perturbation treatment of the electronic coupling.
However, with $V \approx 100$ meV FGR still gives a quite reasonable
value of the carrier mobility
although the electronic coupling is already too large to be formally considered as a perturbation~\cite{Troisi11rev}.
We speculate the success of the hopping picture might be partly ascribed to the cancellation of errors,
i.e., the fast quadratic growth compensates the drawback of 
neglecting the contribution from the coherent transport.
In Figure~\ref{fig:mu-v}b
we have shown the mean free path ($l_{\textrm{mfp}}= v \tau $) of the charge carrier
calculated by TD-DMRG
with the group velocity $v$ and relaxation time $\tau$
estimated by~\cite{Nenad19}:
\begin{equation}
\begin{aligned}
    v & =\sqrt{\braket{\hat j (0) \hat j(0)}} \\
    \tau& =\frac{1}{2} \int_{-\infty}^\infty \abs{\frac{\Re C(t) } {\Re C(0) }} dt .
\end{aligned}
\end{equation}
We find that when $V < 20 \ \textrm{meV}$, $l_{\textrm{mfp}} / R$ is smaller than 1, so a localized hopping picture is well suited for this regime. For the actual parameter of the rubrene crystal ($V=83 \ \textrm{meV}$), $l_{\textrm{mfp}} / R$ is found to be 2.4, which indicates that
neither hopping nor band picture is perfectly suitable
for this regime and an unbiased method like TD-DMRG is necessary.

\begin{figure}
  \includegraphics[width=.48\textwidth]{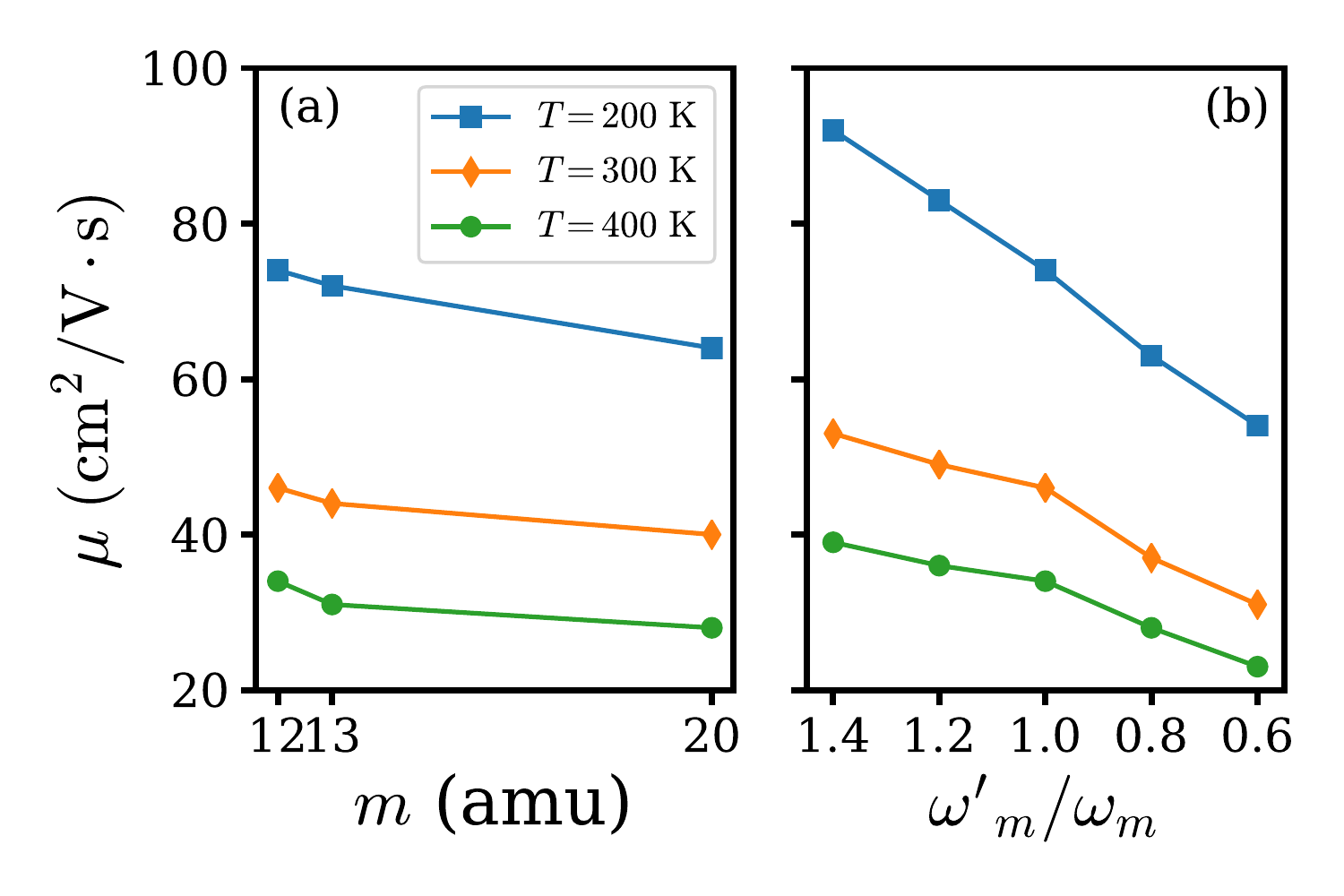}
  \caption{The isotope effect on charge transport in the rubrene crystal
  from $T = 200 \ \textrm{K}$ to $T = 400 \ \textrm{K}$.
  (a) Carrier mobility with increasing carbon mass $m$ and respective $\omega_m$ and $\lambda_m$ are obtained by first-principle calculation.
  (b) Carrier mobility at different vibrational frequencies $\omega'_m$ with
  the original frequencies $\omega_m$ as the unit and $\lambda_m$ held as a constant.}
  \label{fig:mu-isotope}
\end{figure}

The isotope effect for charge transport is controversial in that a number of methods predict
different results.
Here, we use the numerically exact TD-DMRG to study
the isotope effect for charge transport in rubrene crystal.
Our previous work points out that the nuclear tunneling effect
plays an indispensable role in the isotope effect~\cite{Jiang18}.
In this regard, TD-DMRG serves as a suitable tool because
it treats both the electronic and vibrational degrees of freedom 
on an equal footing quantum mechanically.
In this work, we use
two different approaches to simulate isotope substitution
and they eventually lead to the same conclusion.
The first approach is to perform first-principle
quantum chemistry calculation on isotope substituted rubrene molecule
to obtain new sets of parameters for Eq.~\ref{eq:ham}
shown in Figure~\ref{fig:mu-isotope}a.
To preclude the effect of numerical error, we also calculated the mobility
of a hypothetical rubrene where the atomic mass of carbon is 20 amu.
The second approach to simulate the isotope effect is 
to scale all vibration
frequencies in native rubrene by a constant factor while holding the respective
reorganization energy constant
as illustrated in Figure~\ref{fig:mu-isotope}b.
The calculated mobility by the two approaches shows that 
TD-DMRG predicts negative isotope effect, 
which is in agreement with the experimental result~\cite{Ren17}.
The absolute value of isotope effect $(\frac{\mu'}{\mu}-1) \times 100\%$ 
for $^{13}\textrm{C}$ substitution is found to be around 4\% by TD-DMRG.

\begin{figure}
  \includegraphics[width=.48\textwidth]{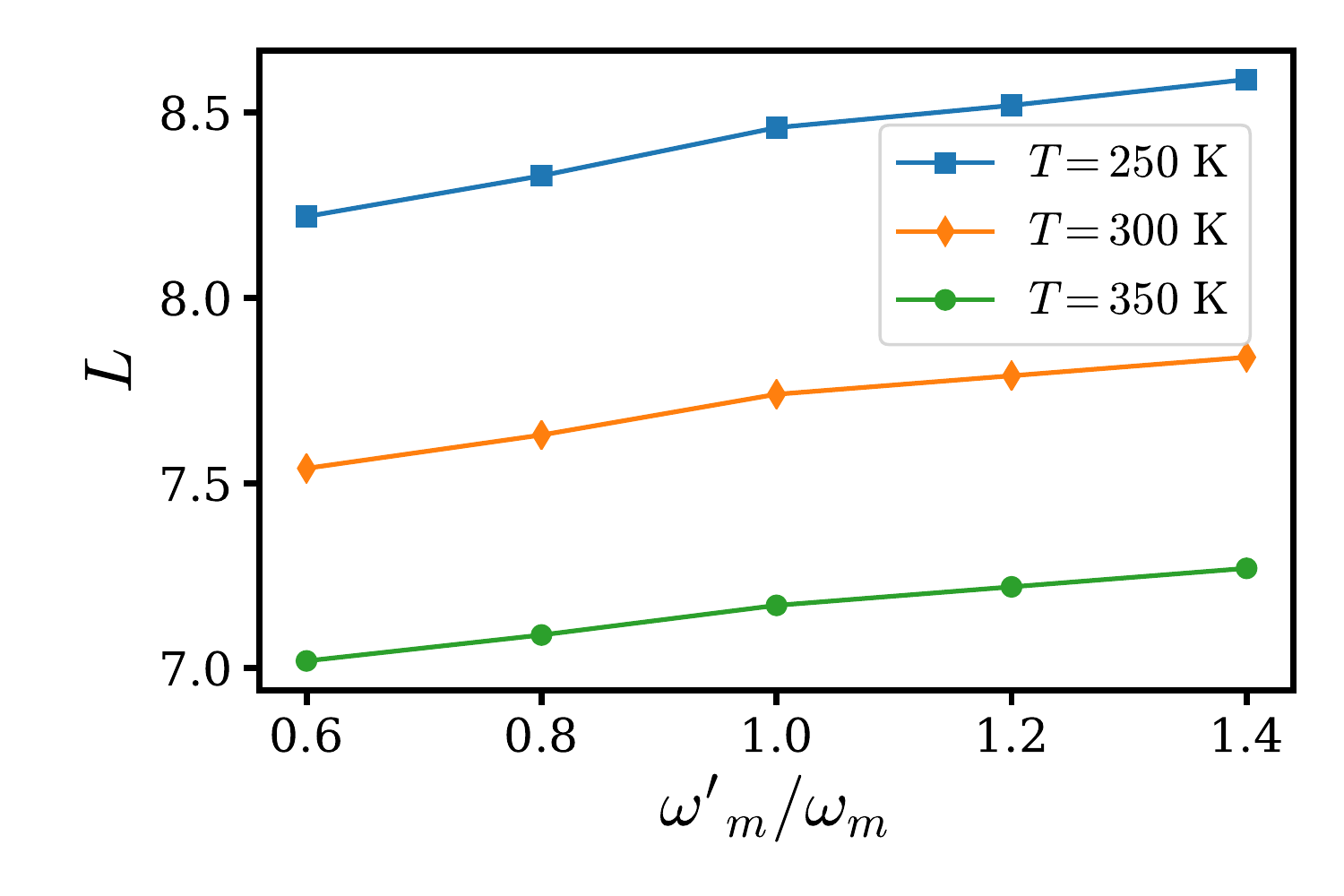}
  \caption{Coherent length $L$ defined in Eq.~\ref{eq:coherent-length} at different
  vibrational frequencies from 250 K to 350 K. $\lambda_m$ for each mode is held as a constant.}
  \label{fig:coherent-length}
\end{figure}

To further analyze the origin of the negative isotope effect from
TD-DMRG point of view, we calculated the coherent length $L$
from the reduced thermal equilibrium density operator 
$\sigma = \textrm{Tr}_{\textrm{ph}}\{\rho_{\beta}\}$
defined as~\cite{meier1997polarons}:
\begin{equation}
\label{eq:coherent-length}
    L=\frac{(\sum_{ij}^N|\sigma_{ij}|)^2}{N\sum_{ij}^N|\sigma_{ij}|^2}.
\end{equation}
For a completely localized thermal equilibrium state, no coherence is present,
so $\abs{\sigma_{ij}} = N^{-1} \delta_{ij}$ and $L=1$.
For completely delocalized thermal equilibrium state, $\abs{\sigma_{ij}} = N^{-1}$
and $L=N$.
So $L$ measures the length scale of charge delocalization
or the polaron size.
The calculated $L$ for different vibrational frequency representing
the effect of isotopic substitution is shown in Figure~\ref{fig:coherent-length}.
For temperature ranging from $T=250 \ \textrm{K}$ to $T=350 \ \textrm{K}$,
$L$ decreases as $\omega'_m/\omega_m$ decreases, which implies that 
heavy isotopic substitution reduces the polaron size.

\begin{figure}
  \includegraphics[width=.48\textwidth]{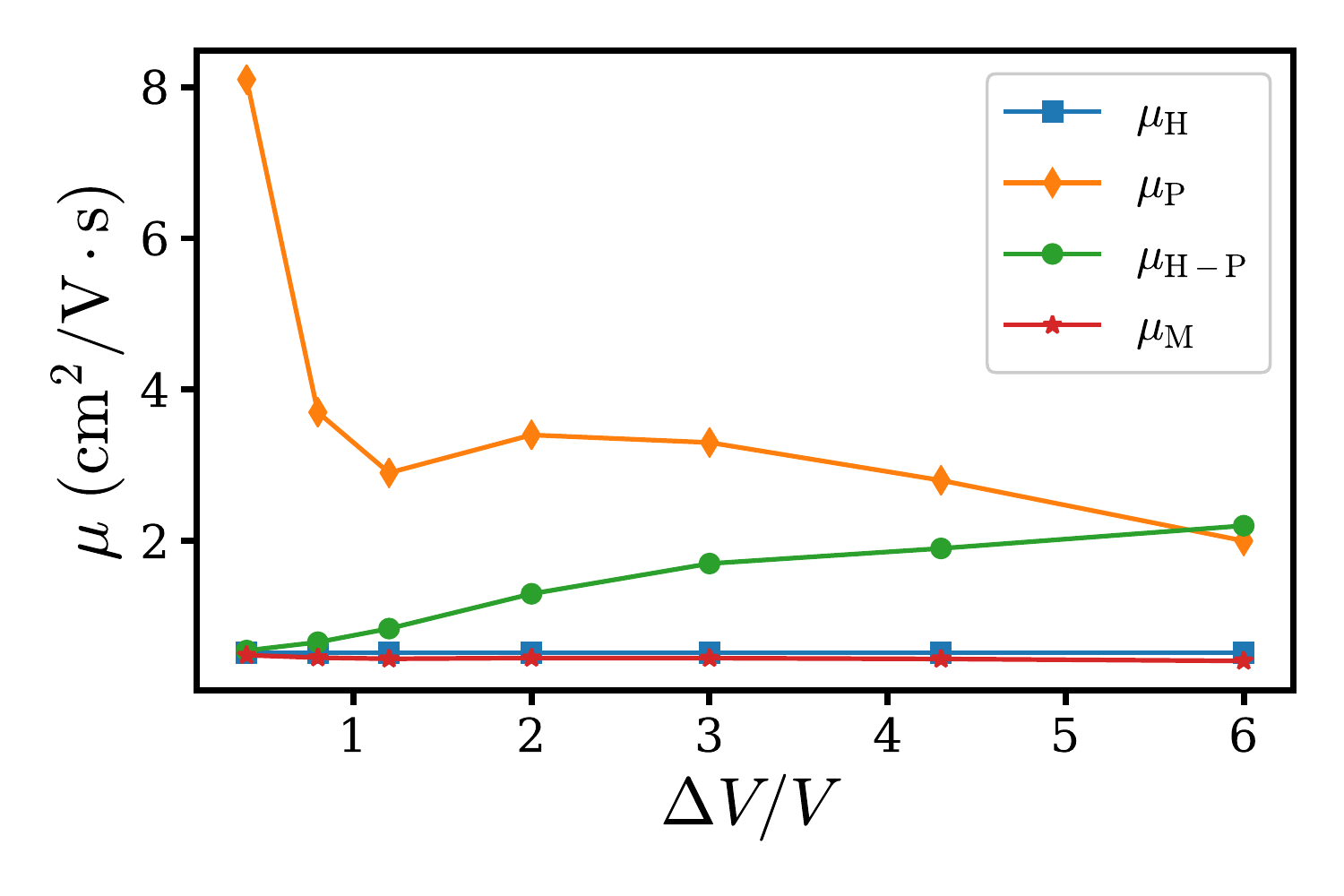}
  \caption{Carrier mobility as a function of the standard deviation of the transfer integral ($\Delta V$) over the mean transfer integral ($V$). $\mu_\textrm{H}$ denotes mobility with only Holstein coupling; $\mu_\textrm{P}$  denotes mobility with only Peierls coupling; $\mu_\textrm{H-P}$  denotes mobility with both Holstein and Peierls coupling; $\mu_\textrm{M}$ is obtained by Matthiessen’s rule $1/ \mu_\textrm{M} =1/\mu_\textrm{H} +1/\mu_\textrm{P}$ .}
  \label{fig:mu-dv}
\end{figure}

\revrevmod{
Finally, we move on to discuss off-diagnoal disorder as it has been proposed as the dominant mechanism for transport, which was not included in the model Hamiltonian Eq.~\ref{eq:ham}. 
This is an interesting mechanism, already proposed by Munn-Silbey in 1980~\cite{Silbey851, Silbey852} and strongly revived recently~\cite{TROISI06, Fratini17}. Our previous study based on molecular dynamics and quantum tunneling enabled hopping model for pentacene indicated that dynamic disorder can indeed limit transport in one-dimensional but not for higher dimension, a conclusion in accordance with Anderson’s theorem~\cite{Wang10Multicale}. 
Thus a comprehensive consideration of off-dynamic disorder should be carried out in high dimension. So far, our TD-DMRG algorithm is structured for 1d case. For 2d, a tensor network type renormalization scheme could be considered but the computational complexity is much higher, deserving further efforts to develop efficient algorithms. But in a preliminary attempt to unravel the off-diagonal term in the present study of 1-d case, we set $V = 8.3 \ \textrm{meV}$, in the hopping limit so that we can compare with our previous case study~\cite{Wang10Multicale}.
The off-diagonal electron-phonon coupling is included in the system via the Holstein-Peierls Hamiltonian similar with Eq.~\ref{eq:ham} except that $\hat H_{e-ph}$ becomes:}
\begin{equation}
    \hat H_{e-ph}  = \sum_{n, m} g^{(1)}_m \omega_m (b^\dagger_{n, m} + b_{n, m}) c^\dagger_n c_n 
     +  \sum_{n, m} g^{(2)}_m \omega_m (b^\dagger_{n, m} + b_{n, m}) (c^\dagger_n c_{n+1} + c^\dagger_{n+1} c_n)
\end{equation}
\revrevmod{
Here we assume $g^{(1)}_m g^{(2)}_m=0$ and only one mode contributes to the Peierls coupling. The parameters of intramolecular vibration are adopted from the 4-mode parameters shown in the supporting information and the frequency of the intermolecular vibration mode is set to 50 $\textrm{cm}^{-1}$.
The mobility calculated by TD-DMRG as a function of the standard deviation of the transfer integral ($\Delta V$) over the mean transfer integral ($V$) 
is shown in Figure~\ref{fig:mu-dv}.
We consider mobilities with three kinds of electron-phonon coupling paradigms in Figure~\ref{fig:mu-dv}: 1) mobility with only Holstein coupling $\mu_\textrm{H}$; 2) mobility with only Peierls coupling $\mu_\textrm{P}$ and 3) mobility with both Holstein and Peierls coupling $\mu_\textrm{H-P}$. With only Holstein coupling there is actually no transfer integral fluctuation and $\mu_\textrm{H}$ is a constant. With only Peierls coupling, $\mu_\textrm{P}$ firstly decreases, and then increases, and finally decreases again. In the absence of Holstein coupling, $\mu_\textrm{P}$ in $\Delta V \rightarrow 0$ limit approaches infinite which is why $\mu_\textrm{P}$ firstly decreases. The subsequent increase and decrease can be ascribed to stronger phonon-assisted current at larger $\Delta V$ and disorder limited charge transport when $\Delta V \rightarrow \infty$ respectively. On the contrary, $\mu_\textrm{H-P}$ steadily grows from $\Delta V /V=0.4$ to $\Delta V /V=8$ due to phonon-assisted current. We also show in Figure~\ref{fig:mu-dv} the mobility obtained by Matthiessen’s rule $1/ \mu_\textrm{M} =1/\mu_\textrm{H} +1/\mu_\textrm{P}$ and we find out that although the rule is applicable at small $\Delta V$ limit, it fails to describe phonon-assisted current which is dominant at large $\Delta V$ regime. Our result is consistent with our previous report~\cite{Wang10Multicale}.}

In summary, we have demonstrated the finite temperature TD-DMRG works as a general and powerful tool to calculate the carrier mobility for organic semiconductors covering a wide range of electron-phonon coupling strength from the hopping to the band regimes. When coupled with molecular parameters from the first-principle calculation, taking rubrene as example, TD-DMRG successfully accounts for the experimentally observed “band-like” transport behavior. The long-standing controversial over the isotope effect on mobility was definitely demonstrated to be negative. We also carefully compared with the well-established solutions at the hopping and band limits and found that TD-DMRG is able to reproduce the analytical result in the hopping limit and approach the asymptotic behavior in the band limit. \revrevmod{At last, we show that TD-DMRG is able to take off-diagonal electron-phonon coupling into consideration as TD-DMRG is quite a general approach. However, a comprehensive description for the off-diagonal disorder should go to higher dimension for more general molecular parameters, an intriguing subject for future work.}

\begin{acknowledgement}
This work is supported by the National Natural Science Foundation of China through the project ``Science CEnter for Luminescence from Molecular Aggregates (SCELMA)'' Grant Number 21788102, as well as by the Ministry of Science and Technology of China through the National Key R\&D Plan Grant Number 2017YFA0204501. J.R. is also supported by the Shuimu Tsinghua Scholar Program.
The authors also gratefully thank Prof. Hua Geng and Dr. Yuqian Jiang for helpful discussions.
\end{acknowledgement}

\begin{suppinfo}
    \begin{itemize}
        \item Filename: si.pdf \\
            Vibrational modes of the rubrene molecule,
            Poincaré recurrence observed in the hopping limit and
            brief overview of TD-DMRG algorithm.
    \end{itemize}
\end{suppinfo}

\bibliography{reference.bib}
\end{document}


\section{Vibrational Modes of Rubrene Molecule}
\begin{figure}[h]
  \includegraphics[width=.48\textwidth]{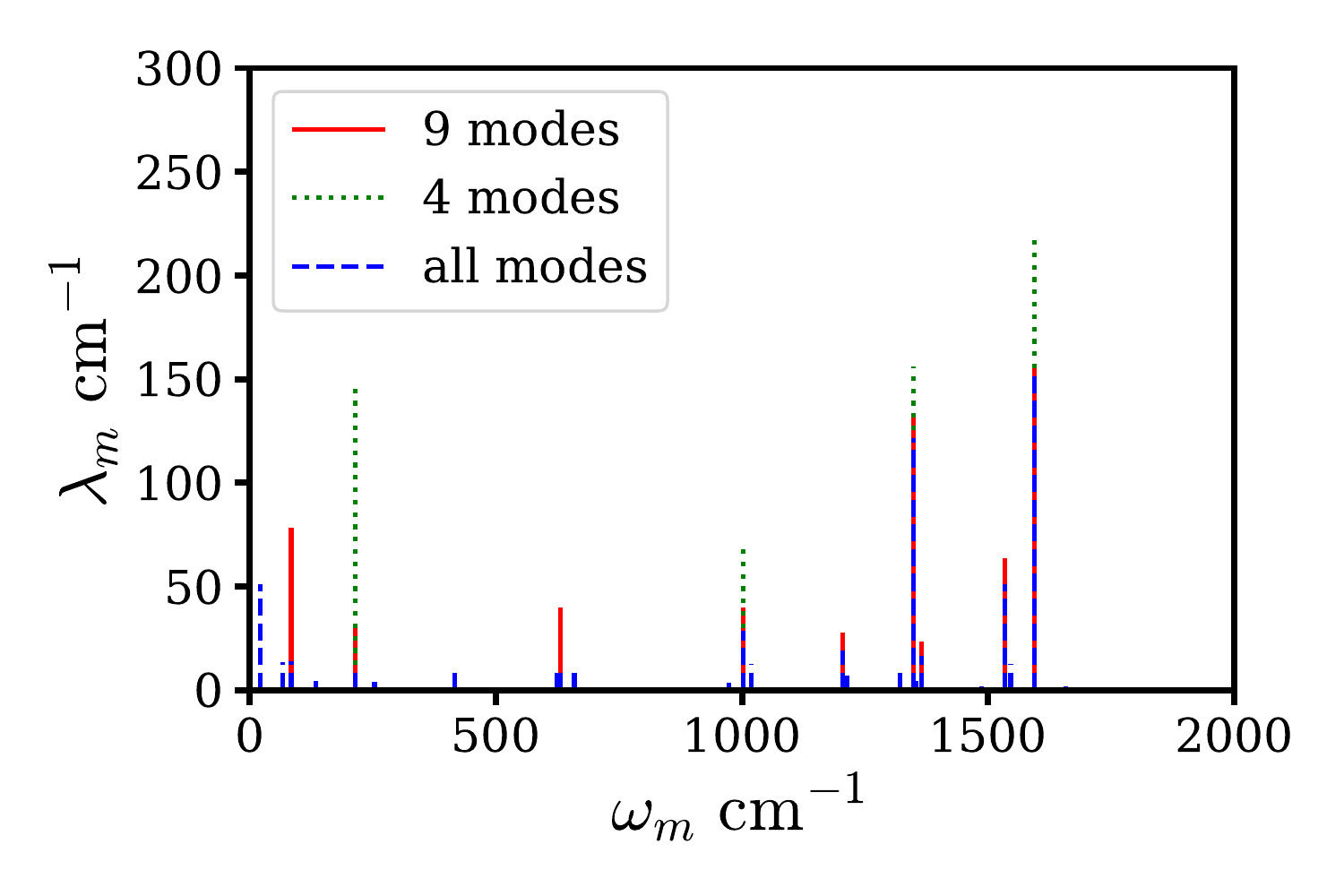}
  \caption{Vibrational frequency $\omega_m$ and corresponding reorganization energy
  $\lambda_m$ for all normal modes of rubrene as well as the 9 vibrational modes used in our calculation and the 4 vibrational modes used to validate our treatment.}
  \label{fig:si-modes}
\end{figure}

The rubrene molecule (\ce{C42H28}) consists of 70 atoms and there are more than 40
vibrational modes that contribute to electron-phonon coupling.
It is virtually impossible for full-quantum numerically exact methods such
as TD-DMRG to include all modes explicitly.
Luckily, it is usually not necessary to treat all modes on equal footing,
since only a few modes contribute to most of the total reorganization energy,
and there exists theoretical evidence that the major charge transport mechanism can be fully understood with one single vibration mode~\cite{Wang15}. 
Thus, a crude approximation is to drop modes with $\lambda_m$
smaller than a certain threshold. However, this approach
may have a notable impact on the total reorganization energy and
further the calculated carrier mobility.
In this work we improve the approach by adding $\lambda_m$ of the dropped mode to the closest retained mode.
In this way, the total reorganization energy is conserved.
For data reported in the main text, we have dropped modes that
satisfies $\lambda_m < 20 \ \textrm{cm}^{-1}$
which reduces the total number of modes for each molecule to 9.
In Figure~\ref{fig:si-modes} we show
all vibrational modes obtained by first-principle calculation
and the actual 9 modes used in TD-DMRG simulation.
To validate the choice of the modes, we have also tested
carrying out TD-DMRG calculation with 4 vibrational modes obtained by
dropping modes with $\lambda_m < 50 \ \textrm{cm}^{-1}$,
also shown in Figure~\ref{fig:si-modes}.
As outlined in Table~\ref{tb:mobility-modes}, the absolute value
as well as the tendency of carrier mobility does not vary much
between 9-mode result and 4-mode result,
implying that using the 9 modes reported in this work is able to catch
the essential feature of the system.
A practical concern preventing us from using even less modes is that
with less modes the correlation function $C(t)$ can not decay to zero, in which case a phenomenological broadening parameter might have
to be applied.
Lastly we note that there exist several vibrational modes that possess relatively
low frequency ($\omega_m < 25 \ \textrm{cm}^{-1}$), which are numerically
infeasible for TD-DMRG at finite temperature since a large number of harmonic oscillator eigenbasis is required
to represent each low-frequency mode. 
So in our 9-mode calculation
the frequency is increased to $84 \ \textrm{cm}^{-1}$, 
as shown in Figure~\ref{fig:si-modes}.

\begin{table}[h]
\caption{Carrier mobility in $\textrm{cm}^2 / \textrm{V} \cdot \textrm{s}$ with 9 vibrational modes and 4 vibrational modes at temperature from $200 \ \textrm{K}$ to $400 \ \textrm{K}$.}
\label{tb:mobility-modes}
\centering
\begin{tabular}{c c c c}
\hline
$T$ (K)    & 200  & 300 &  400     \\ \hline
9 modes & 66 & 47 & 34   \\ 
4 modes & 59 & 40 & 30   \\ \hline
\end{tabular}
\end{table}

For $^{13}\textrm{C}$ and $^{20}\textrm{C}$ substituted rubrene,
quantum chemistry calculation is performed by the Gaussian 09 package
with the B3LYP functional and 6-31G$^*$ basis set~\cite{G09}.

\section{Poincaré Recurrence in the Hopping Limit}
\begin{figure}
  \includegraphics[width=.48\textwidth]{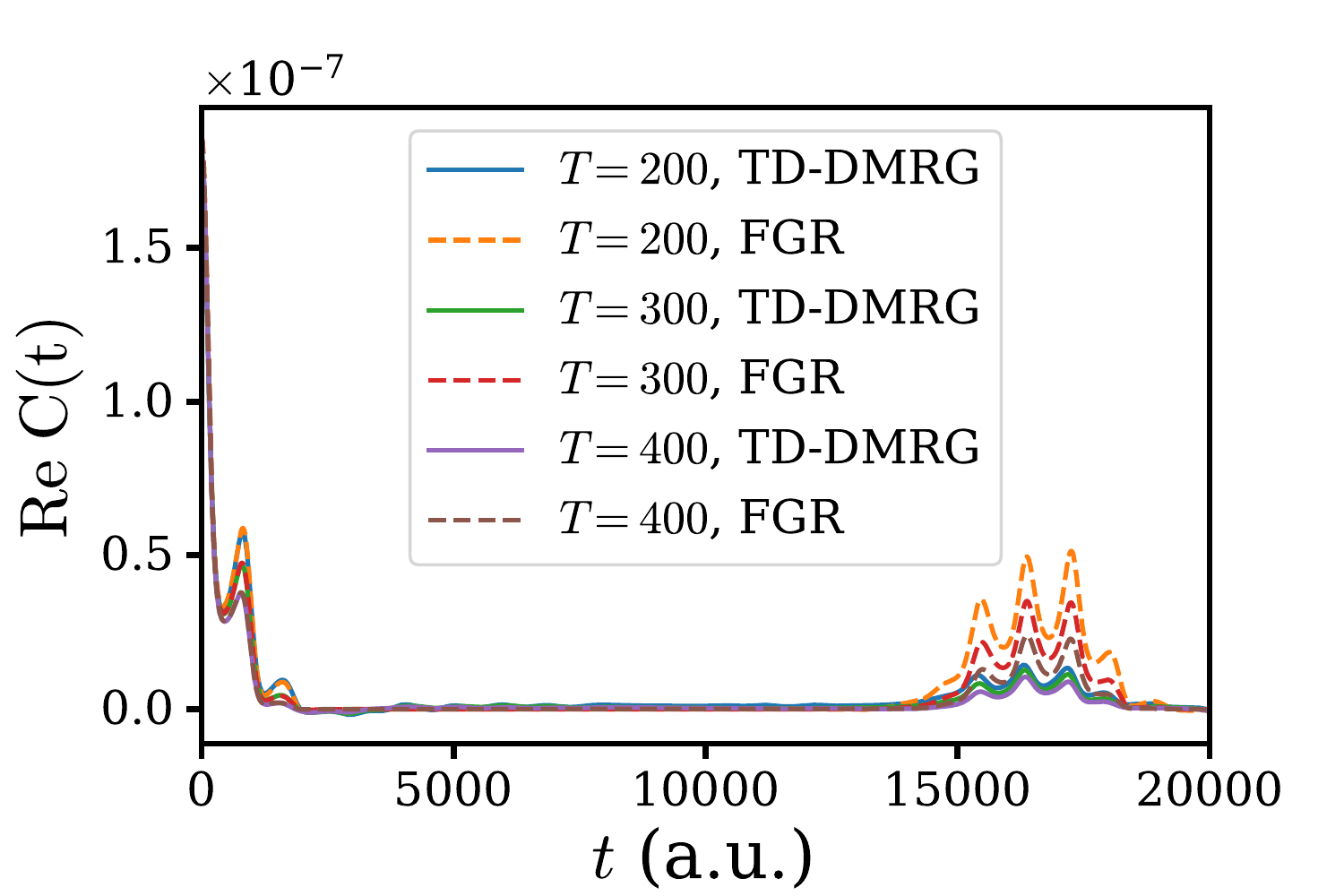}
  \caption{Correlation function $C(t)$ in terms of atomic unit obtained by TD-DMRG (solid) and FGR (dashed) in the hopping limit ($V=8.3$ meV).}
  \label{fig:si-recur}
\end{figure}

In the hopping limit ($V=8.3$ meV) Poincaré recurrence is observed after around 15000 a.u. as showin in Figure~\ref{fig:si-recur}.
With time evolution parameters used in the main text, the amplitude of the recurrence predicted by TD-DMRG is smaller than the amplitude predicted by FGR. Possible reasons include the inaccuracy of the perturbation approximation and the numeric error accumulated with TD-DMRG time evolution.
The recurrence is unlikely to happen for realistic materials due to the presence of various dissipations. When integrating the correlation function the integration time limit is set to be smaller than the recurrence time to exclude the effect of the recurrence.

\section{Brief Overview of the TD-DMRG Algorithm}
The TD-DMRG algorithm in the language of matrix product state (MPS)
and matrix product operator (MPO) has already been reviewed in detail~\cite{Schol11, Scholl19}. 
Here we briefly overview the basic idea behind TD-DMRG and discuss about
the major factors that are relevant to the efficiency of finite temperature TD-DMRG simulations.
The MPS ansatz describes the system wavefunction as the product of a matrix chain:
\begin{equation}
\label{eq:mps}
    \ket{\Psi}  = \sum_{\{a\},\{\sigma\}}
     A^{\sigma_1}_{a_1} A^{\sigma_2}_{a_1a_2} \cdots
           A^{\sigma_N}_{a_{N-1}}  \ket{ \sigma_1\sigma_2\cdots\sigma_N }
\tag{S1}
\end{equation}
$A^{\sigma_i}_{a_{i-1}a_{i}}$ are matrices in the chain connected by indices $a_i$. $\{ \cdot \}$ in the summation represents the contraction of the respective connected indices, 
and $N$ is the total number of degrees of freedom (DOFs) in the system.
The dimension of $a_i$ is called (virtual) bond dimension,
while the dimension of $\sigma_i$ is called physical bond dimension.
In principle, TD-DMRG is able to solve the time dependent Schr\"odinger equation in an exact manner if the bond dimension is infinite. In practice, the accuracy of the method can be systematically improved by using a larger bond dimension, until convergence of interested physical observables within arbitrary convergence criteria. Therefore, the method is sometimes described as “nearly exact”.

The efficiency of finite temperature TD-DMRG is mostly controlled by four factors:
1) The number of DOFs in the system $N$, 
2) the (virtual) bond dimension of the MPS ansatz, 
3) the physical bond dimension (the dimension of $\sigma_i$) and 
4) the energy scale (relative to evolution time) of the system Hamiltonian. More specifically:
\begin{enumerate}
    \item Systems with more DOFs require more matrices to represent. In practice, TD-DMRG for vibronic systems is able to handle at most 1000 DOFs. 
    \item The (virtual) bond dimension controls the number of parameters for the MPS ansatz. The higher bond dimension, the more accurate MPS representation of the system wavefunction. The bond dimension required for numerically exact calculation varies dramtically with systems. For systems with very strong correlation, the bond dimension could be more than 10000, whereas for vibronic systems bond dimension within 100 is usually sufficient.
    \item The physical bond dimension is determined by the basis required for each DOF. For vibration DOFs the sizes of the harmonic oscillator eigenbasis are theoretically infinite and for numerical calculation it is usually truncated to a finite value.
    \item The energy scale of the Hamiltonian affects the time integration step for the time evolution. If the energy scale becomes larger, the wavefunction coefficients and the value of physical observables of the system typically evolves in a faster manner and thus smaller time integration step is required. 
\end{enumerate}

\bibliography{reference.bib}